\documentclass[aps,prl,twocolumn,reprint,superscriptaddress]{revtex4}
\usepackage{graphicx}
\usepackage{amsmath}
\usepackage{amssymb}
\usepackage{colordvi}
\usepackage{mathrsfs}
\usepackage{bm}
\usepackage{verbatim}
\usepackage{dcolumn}
\usepackage{epsfig}
\usepackage{subfigure}
\usepackage[colorlinks,allcolors=blue]{hyperref}
\usepackage{ulem}
\usepackage{dsfont}
\usepackage{makecell}
\usepackage{lipsum,epsfig,dsfont}
\begin{document}
\title{Braiding Induced by Finite-Size Effect in One-Dimensional Topological Superconductors}
\author{Hongfa Pan}
\affiliation{International Centre for Quantum Design of Functional Materials, CAS Key Laboratory of Strongly-Coupled Quantum Matter Physics, and Department of Physics, University of Science and Technology of China, Hefei, Anhui 230026, China}
\author{Zhengtian Li}
\affiliation{International Centre for Quantum Design of Functional Materials, CAS Key Laboratory of Strongly-Coupled Quantum Matter Physics, and Department of Physics, University of Science and Technology of China, Hefei, Anhui 230026, China}
\author{Jinxiong Jia}
\affiliation{International Centre for Quantum Design of Functional Materials, CAS Key Laboratory of Strongly-Coupled Quantum Matter Physics, and Department of Physics, University of Science and Technology of China, Hefei, Anhui 230026, China}
\author{Zhenhua Qiao}
\email[Correspondence author:~~]{qiao@ustc.edu.cn}
\affiliation{International Centre for Quantum Design of Functional Materials, CAS Key Laboratory of Strongly-Coupled Quantum Matter Physics, and Department of Physics, University of Science and Technology of China, Hefei, Anhui 230026, China}
\affiliation{Hefei National Laboratory, University of Science and Technology of China, Hefei 230088, China}
\date{\today}

\begin{abstract}
  We investigate the transport properties of Majorana zero mode (MZM) and Majorana Kramers pair (MKP) in one-dimensional topological superconductors, respectively. An effective model is established for braiding of MZMs and MKPs. We employ the $d_{x^{2}-y^{2}}$-wave topological superconductors to embody the effective model for braiding of MKPs by utilizing finite-size effects and locally tunable coupling parameters. We show how to construct the state initialization and readout via gate control. We also use this method for braiding MZMs in s-wave topological superconductors. Our proposal presents a promising avenue for experimentally verifying the non-Abelian statistical properties of MZMs and MKPs, with implications for topological quantum computing.
\end{abstract}

\maketitle
\textit{Introduction--.} Recently, topological superconductors (TSCs) have attracted extensive research interest due to their promising for fault-tolerant topological quantum computation~\cite{C. Nayak2008,Sato2017,Alicea2012,Kitaev2001,D. Aasen2016,Liu Xiong-Jun2020}. As quasi-particles in TSCs respectively without and with time-reversal symmetry, both Majorana zero mode (MZM) and Majorana Kramers pair (MKP) can exist at the ends of one-dimensional (1D) TSCs~\cite{Kitaev2001,R. M. Lutchyn2010,Y. Oreg2010,C. L. M. Wong2012} or at the cores of 2D topological superconducting vortices~\cite{M. Sato2009,J. D. Sau2010,L. Fu2008,Qi2009,Green2000}, exhibit non-Abelian statistics, and can be used to construct quantum bits~\cite{C. Nayak2008,Alicea2012,D. Aasen2016,X.-J. Liu2014}. However, so far, most experimental signals such as zero-bias conductance peaks~\cite{K. T. Law2009,Chun-Xiao Liu2018,J. Liu2012,C. Reeg2018}, cannot provide convincing evidence of MZMs or MKPs. A direct confirmation of the presence of MZMs or MKPs requires the braiding experiment~\cite{C. Nayak2008,Alicea2011,X.-J. Liu2014}, that involves the exchange of MZMs or MKPs to confirm the non-Abelian statistics. The experimentally viable braiding scheme is fraught with numerous complexities ~\cite{Alicea2011,Alicea2012,X.-J. Liu2014,P. Gao2016,Y. Tanaka2022,J. Manousakis2020,C. Schrade2018} and remains further exploration.

As an ideal platform, 1D TSCs can be realized via superconductor proximity effect in various 1D systems~\cite{R. M. Lutchyn2010,F. Zhang2013,F. Pientka2017,Y. Oreg2010,E. Gaidamauskas2014}. On one hand, in the absence of time-reversal symmetry, MZMs at ends of a 1D TSC~\cite{R. M. Lutchyn2010,Y. Oreg2010,F. Pientka2017} can give rise to nonlocal Fermion states involving both occupied and unoccupied states. The exchange of two non-paired MZMs can facilitate the transition between different states~\cite{C. Nayak2008,Amorim2015,Sanno2021,W. Chen2022,J. Liu2021,Y.-F. Zhou2019}. On the other hand, in the presence of time-reversal symmetry, the magnetic field is avoided, mitigating the deleterious effects of field-induced suppression of superconducting energy gap. MKPs emerge at different ends of a 1D TSC~\cite{F. Zhang2013,E. Gaidamauskas2014}, forming four possible nonlocal Fermion states~\cite{X.-J. Liu2014,Y. Tanaka2022,Schrade2022}. After braiding, it can realize a transition between the states with same Fermion parity~\cite{X.-J. Liu2014}. As a consequence, the ability to experimentally verify topological superconductivity, as well as to pursue topological quantum computing, is significantly enhanced by designing experimentally viable braiding protocols.

In this Letter, inspired by the interplay between the quantum dot and semiconductor nanowires hosting MZMs~\cite{M. T. Deng2016,M.-T.
Deng2018,Corneliu Malciu2018}, we theoretically explore the transport schemes of MZMs and MKPs based on the effective models. As concrete examples, the system models are constructed to include two parts, i.e., one isolated MZM (MKP) and two coupled MZMs (MKPs). By adiabatically tune the coupling strength between the two parts, we theoretically realize the directional transport of MZMs (MKPs).

By employing the transport mechanism introduced in Ref.~[\onlinecite{J. Liu2021}], we formulate an effective braiding model of a quasi-T-junction structure to facilitate the braiding of MKPs. In principle, numerous viable schemes can be proposed to realize the braiding of MKPs in specific materials within the framework of effective models. To be specific, we utilize the finite-size effect to implement the braiding of MKPs within the $d_{x^{2}-y^{2}}$-wave TSC~\cite{C. L. M. Wong2012,Y. Tanaka2022}. Our approach, requiring the local adjustments of coupling parameters, is completely from the conventional ones of globally tuning the chemical potentials and other parameters to reach the movement of topological domain walls, thereby greatly enhancing its experimental feasibility. In addition, the initialization of state is crucial for the result measurement in the experiment, which was usually ignored. We design a gate-controlled proposal to determine the initial state. After braiding, we reverse the initialization operation to realize the local measurement. This method can be used in time-reversal symmetry breaking system. Similarly, we also demonstrate the braiding of MZMs by utilizing the 1D s-wave TSC nanowires~\cite{R. M. Lutchyn2010,Y. Oreg2010}.
\begin{figure}
  \centering
  \includegraphics[width=8.5cm,angle=0]{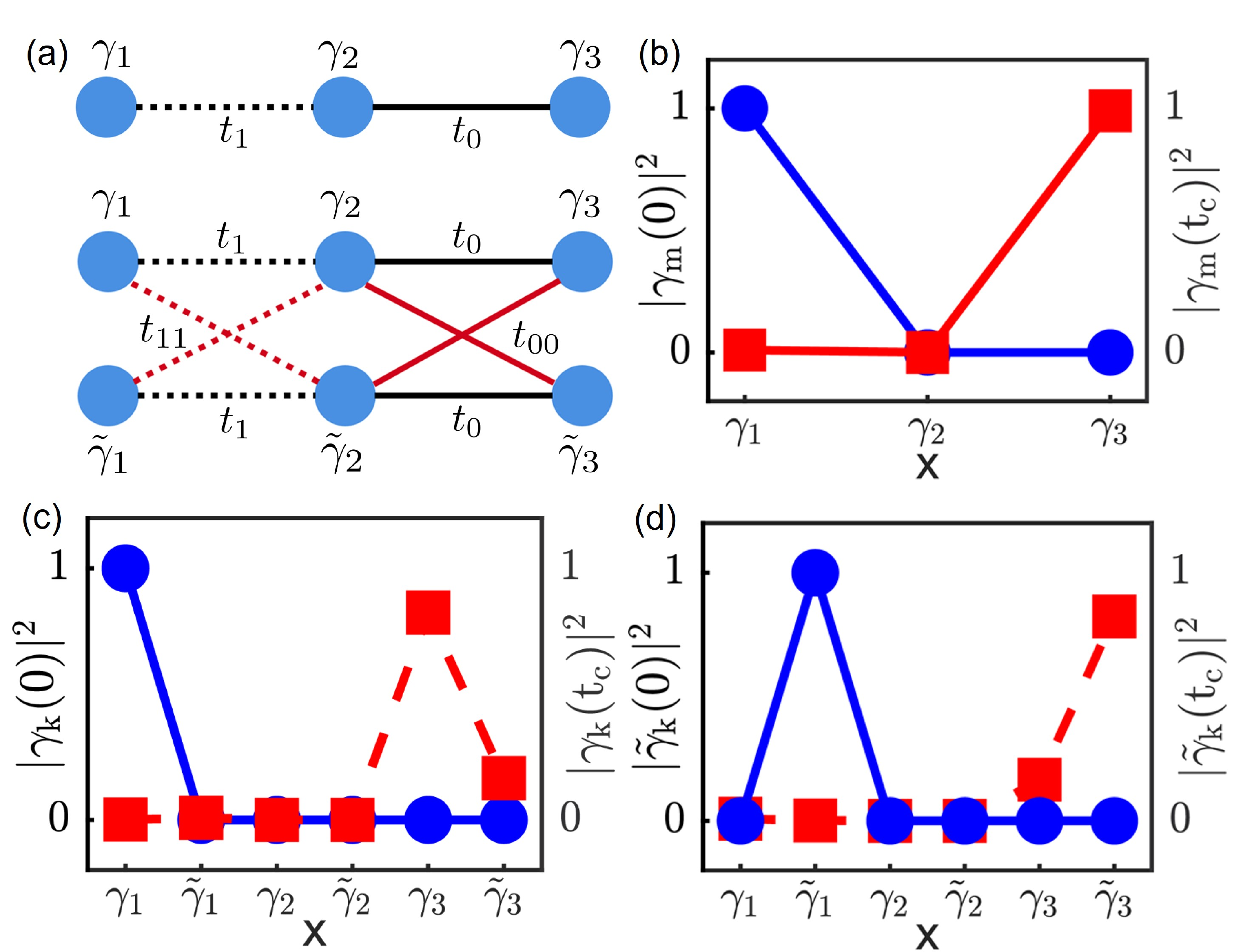}
  \caption{(a) Transport schemes of MZMs (upper) and MKPs (lower). Blue dots represent MZMs and MKPs. (b) Spatial distribution of MZMs. It is denoted in blue (red), when $t_{1}=0$ ($t_{1}=t_{c}$). (c) and (d) Spatial distribution of MKPs. As $t_{1}=t_{11}=0$, spatial distribution is denoted by the blue line. It is denoted by the red line when $t_{1}=t_{11}=t_{c}$.}
		\label{fig1}.
\end{figure}

\textit{Transport of MZMs and MKPs--.} The effective scheme of transport of MZMs is shown in Fig.~\ref{fig1}(a) (see upper panel), where the blue dots denote three MZMs, i.e., $\gamma_{1,2,3}$. The couplings between $\gamma_{1}$($\gamma_{2}$) and $\gamma_{2}$($\gamma_{3}$) are described by $t_{1}$ and $t_{0}$, respectively, leading to the Hamiltonian~\cite{Corneliu Malciu2018}
\begin{eqnarray}
H_{1}=it_{1}\gamma_{1}\gamma_{2}+it_{0}\gamma_{2}\gamma_{3}.
\end{eqnarray}
The coupling $t_0$ is set to be a constant. Consequently, a zero-energy state corresponding to a MZM can persist~\cite{SuppMat}. As shown in Fig.~\ref{fig1}(b), at the initial state, setting $t_{1}=0$ results in the emergence of MZM $\gamma_{1}$; As adiabatically increasing $t_{1}$ to be larger than $t_{0}$, the MZM undergoes a transition to $\gamma_{3}$. As for TSCs with time-reversal symmetry, which can be decoupled into two distinct sectors, it is effectively equivalent to the scenario without time-reversal symmetry, as shown in Fig.~\ref{fig1}(a) (upper panel), in the absence of the couplings denoted by red lines in Fig.~\ref{fig1}(a) (lower panel). Similarly, setting $t_{1}=0$, a MKP formed by $\gamma_{1}$ and $\widetilde{\gamma}_{1}$ emerges. As adiabatically increasing $t_{1}$ to be lager than $t_{0}$, the MKP evolves into the state formed by $\gamma_{3}$ and $\widetilde{\gamma}_{3}$ naturally. Generally, TSCs with time-reversal symmetry cannot be decoupled into two parts. The effective scheme of transport of MKP is shown in Fig.~\ref{fig1}(a) (lower panel). The Hamiltonian reads:
\begin{eqnarray}
H_{2}&=&it_{1}(\gamma_{1}\gamma_{2}+\widetilde{\gamma}_{1}\widetilde{\gamma}_{2})+it_{11}(\gamma_{1}\widetilde{\gamma}_{2}+\gamma_{2}\widetilde{\gamma}_{1}) \nonumber \\
&+&it_{0}(\gamma_{2}\gamma_{3}+\widetilde{\gamma}_{2}\widetilde{\gamma}_{3})+it_{00}(\gamma_{2}\widetilde{\gamma}_{3}+\gamma_{3}\widetilde{\gamma}_{2}).
\end{eqnarray}
The couplings $t_0$ and $t_{00}$ are set to be constant. Consequently, two zero-energy states corresponding to MKPs can persist~\cite{SuppMat}. Setting $t_{1}=0$ and $t_{11}=0$, a MKP formed by $\gamma_{1}$ and $ \widetilde{\gamma_{1}}$ emerges as shown in Fig.~\ref{fig1}(c). As both $t_{1}$ and $t_{11}$ are adiabatically increased to be respectively larger than $t_{0}$ and $t_{00}$, the MKP evolves into the right end as shown in Fig.~\ref{fig1}(d). Therefore, the transport of MZMs or MKPs can be realized by tuning the local coupling parameters.

\begin{figure}
	\centering
  \includegraphics[width=8.5cm,angle=0]{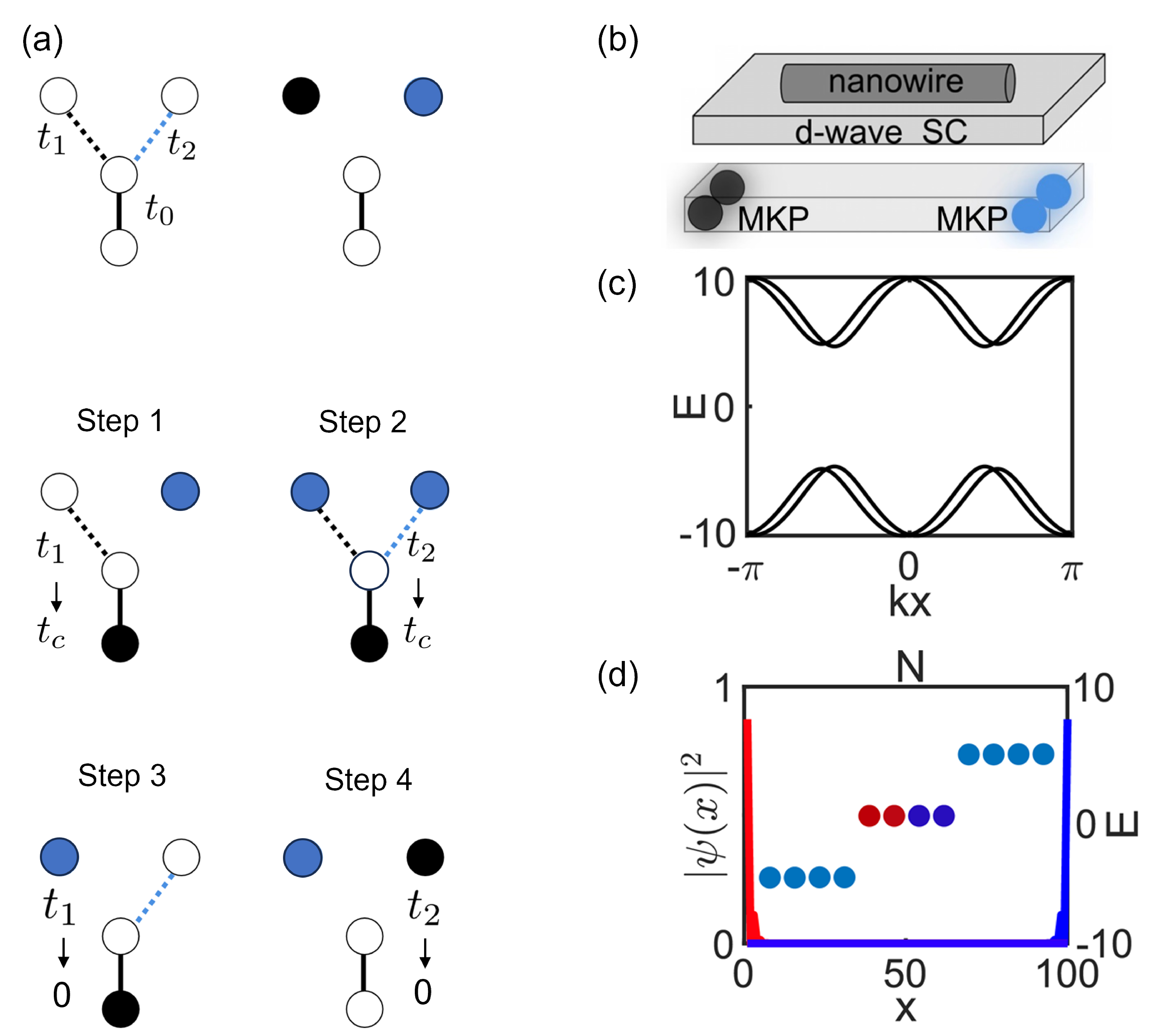}
	\caption{(a) Effective scheme and specific steps for braiding of MKPs. Circles denote MKPs. Black and blue circles denote the sites of two decoupled MKPs in the four steps. (b) A $d_{x^{2}-y^{2}}$ paring TSC nanowire. (c) Energy band of the nanowire in topological regime with a global gap. (d) Four zero-energy states corresponding to two MKPs and their spatial distribution.}
	\label{fig2}
\end{figure}

\textit{Braiding of MKPs--.} The exchange of different MKPs cannot be well defined in one dimension, therefore it is usually necessary to design a T-junction to realize the braiding~\cite{Alicea2011}. Based on the transport method introduced in Ref.~[\onlinecite{J. Liu2021}], we can design the scheme of braiding as shown in Fig.~\ref{fig2}(a). The four circles represent four MKPs with different coupling parameters as shown Fig.~\ref{fig2}(a). The coupling strength $t_{0}$ denoted by black solid line is fixed during the braiding. Initially, two decoupled MKPs emerge as $t_{1}=0$ and $t_{2}=0$. At the first step, a MKP denoted by the black circle evolves as in the transport model by increasing $t_{1}$ to a critical value $t_{c}$. At the second step, increasing $t_{2}$ to $t_{c}$, the other MKP denoted by the blue circle evolves. At the third step, decreasing $t_{1}$ to be $0$, the MKP denoted by the blue circle evolves into the site where the MKP denoted by the black circle was initially located~\cite{SuppMat}. At the last step, decreasing $t_{2}$ to be $0$, the MKP denoted by the black circle evolves into the site where the MKP denoted by the blue circle was initially located. This realizes the exchange of two MKPs. By repeating these four steps of adiabatic manipulation, it can reach two exchanges of the two MKPs. The evolution of the wave function in the braiding can be numerically evaluated by $\left|\gamma(t)\right\rangle=U(t)\left|\gamma(0)\right\rangle$, where the time-evolution operation is $U(t)=\hat{T}{\exp}^{i\int_{0}^{t}d\tau H(\tau)}$ with the time-ordering operator $\hat{T}$ and the Hamilton $H$.

\begin{figure*}
	\centering
	\includegraphics[width=0.9\textwidth]{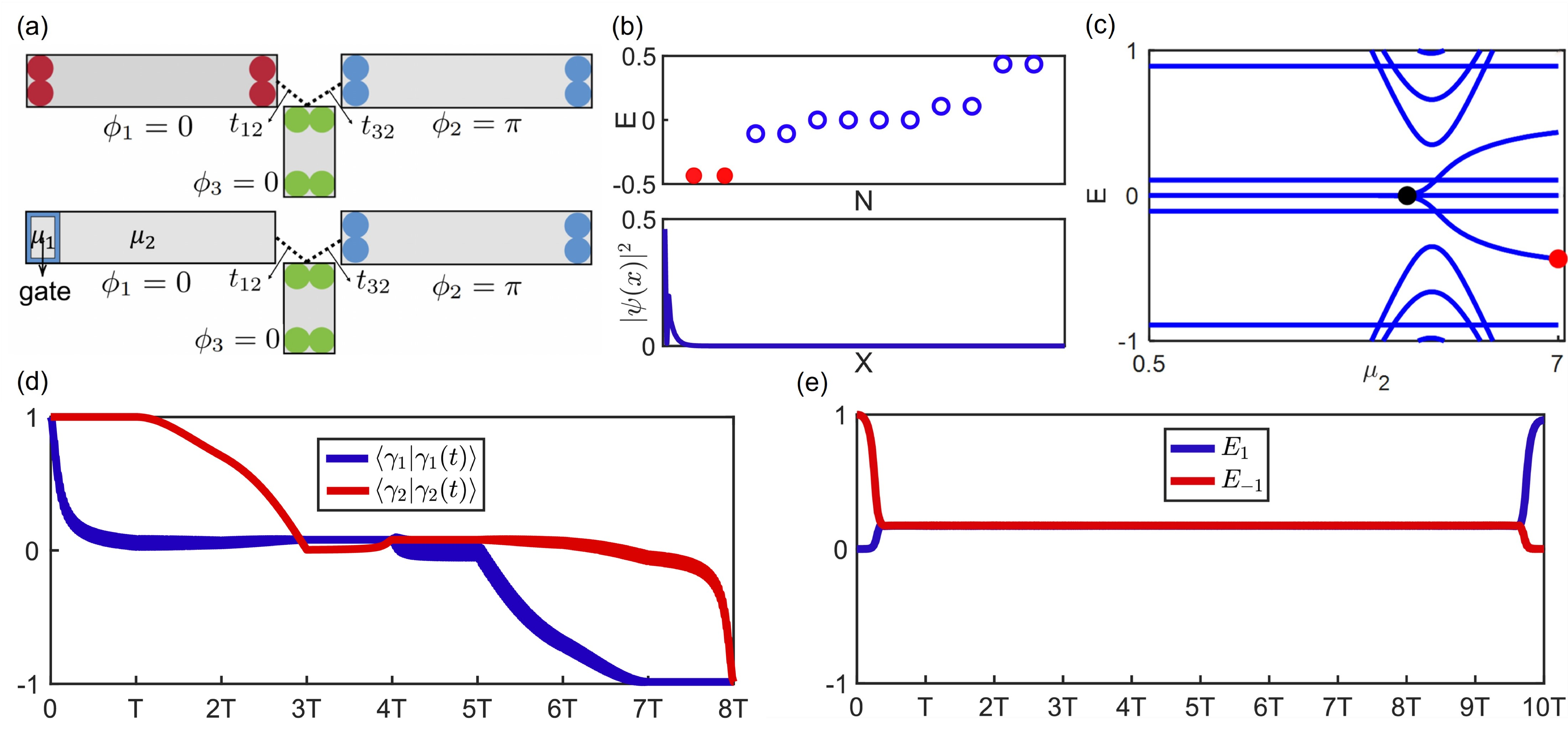}
	\caption{(a) Scheme of braiding in 1D $d_{x^{2}-y^{2}}$ pairing TSCs (upper panel) and scheme considering the initialization (lower panel). (b) Energy distribution and wavefunction distribution of the initial ground state. Red dots denote the state $|0\tilde{0}\rangle$. (c) Energy spectra as a function of $\mu_{2}$. Initial state denoted by red dots evolves into MKPs along with the decrease of $\mu_{2}$. (d) Evolution of MKPs. (e) Evolution considering the initial states. $E_{-1}$ denotes $|0\rangle$, $E_{1}$ denotes $|1\rangle$. After the full operation, state $|0\tilde{0}\rangle$ changes to $|1\tilde{1}\rangle$. $t_{c}=1.2$ and $T=5000$.}
	\label{fig3}
\end{figure*}

As displayed in Fig.~\ref{fig2}(b), a nanowire with $d_{x^{2}-y^{2}}$ paring and Rashba spin-orbit interaction can host MKPs at its ends of the topological region. The corresponding Hamiltonian is written as~\cite{C. L. M. Wong2012,Y. Tanaka2022}
\begin{equation}
	\begin{aligned}
		H_{1 \mathrm{D}} & =-\mu \sum_{j}  c_{j}^{\dagger} c_{j}- t\sum_{j}  c_{j+1}^{\dagger} c_{j}+H_{\mathrm{SO}}+H_{\mathrm{SC}}, \\
		H_{\mathrm{SO}} & =-\frac{i}{2} \alpha_R \sum_{j, \alpha, \beta} c_{j+1, \alpha}^{\dagger} \sigma_y^{\alpha, \beta} c_{j, \beta}+ h.c., \\
		H_{\mathrm{SC}} & =\frac{1}{2} \Delta_0 \sum_j \left(c_{j+1, \uparrow}^{\dagger} c_{j, \downarrow}^{\dagger}-c_{j+1, \downarrow}^{\dagger}c_{j, \uparrow}^{\dagger}\right)+h.c., \\
	\end{aligned}
\end{equation}
where $c^{\dagger}_{j}$ ($c_{j}$) creates (annihilates) an electron at site j, $\alpha$ and $\beta$ are the spin indices, $t$ is the nearest
hopping amplitude. The first and second terms are respectively the chemical potential and the kinetic energy; the third term is the spin-orbit coupling with $\alpha_{R}$ being the coupling strength, and the last term describes the superconducting proximity effect with $\Delta$
being the superconducting pairing amplitude. By setting the length of the nanowire to be $N=100$, $t=1$, $\mu=0$, $\alpha=5$ and $\Delta=10$, the nanowire is in the topological regime with time-reversal symmetry~\cite{Y. Tanaka2022}, where a global band gap opens to host four zero-energy states, as shown in Figs.~\ref{fig2}(c) and \ref{fig2}(d). The four states are located at the two ends of nanowire, forming two MKPs by two states at the same ends related by time-reversal symmetry.

So far, there are many theoretical proposals to braid MKPs based on the effective model. In below, we show how to use finite-size effect to realize the braiding as displayed in Fig.~\ref{fig3}(a). $t_{0}$ can be effectively realized by a short nanowire (i.e., $N_{3}=6$) due to the finite-size effect, which can also be realized effectively by making a small region of a long nanowire be topologically non-trivial. The Hamiltonian of the left (right) long nanowire is denoted by $H_{1t}$ ($H_{3t}$), and that of the short nanowire is denoted by $H_{2t}$. To map the effective model, the phases in different parts are set to be $0$ in the left long nanowire and the short nanowire, and $\pi$ in the right long nanowire respectively, avoiding to introduce extra couplings in the effective model. The Hamiltonian describing the coupling between the left long nanowire and the short nanowire is
\begin{equation}
	H_{12}=-t_{12}\sum_{\alpha}c_{1N, \alpha}^{\dagger}c_{21, \alpha}+h.c..
\end{equation}
The Hamiltonian describing the coupling between the right long nanowire and the short nanowire is
\begin{equation}
	H_{32}=-t_{32}\sum_{\alpha}e^{-i\phi_{3}/2}c_{31, \alpha}^{\dagger}c_{21, \alpha}+h.c.,
\end{equation}
where $c^{\dagger}_{1N,\alpha}$ ($c_{1N,\alpha}$), $c^{\dagger}_{21, \alpha}$ ($c_{21,\alpha}$) and $c^{\dagger}_{31, \alpha}$ ($c_{31, \alpha}$) are creation (annihilation) operators of electrons with spin $\alpha$ at the ends of nanowires 1, 2, and 3, respectively. The total Hamiltonian $H_{T0}=H_{1t}+H_{2t}+H_{3t}+H_{12}+H_{32}$.

The braiding contains four steps. Then, repeating the four steps to realize the exchange of MKPs twice. After braiding, $\gamma_{1}$ and $\gamma_{2}$ accumulate respectively a phase of $\pi$~\cite{SuppMat}. The same result can be obtained for their time-reversal counterparts $\widetilde{\gamma}_{1}$ and $\widetilde{\gamma}_{2}$ due to the time-reversal symmetry. As a consequence, the braiding of MKPs can be realized as displayed in Fig.~\ref{fig3}(d).

\textit{Initialization--.} To test the non-Abelian statistics properties of MKPs, the initialization of state is crucial, and was overlooked previously. Here, we design a small gate to initialize the state of MKPs as displayed in Fig.~\ref{fig3}(a). First, we set the left nanowire to be a topologically trivial state with $\mu_{2}=7$; Then, by applying a small gate, we tune a small region at the left end of the nanowire to be topologically nontrivial with $\mu_{1}=0.5$. Due to the finite-size effect, the bound states deviating from zero energy appear inside the energy gap and are located at the left end of the nanowire as displayed in Fig.~\ref{fig3}(b). The ground state of the system is $|0\tilde{0}\rangle$. By adiabatically controlling the remaining part of the nanowire to be topologically nontrivial by decreasing $\mu_{2}$ from $7$ to $0.5$, we can obtain the initial state, which we need in the braiding as shown in the energy spectra for the T-junction configuration as a function of the chemical potential in Fig.~\ref{fig3}(c). The state denoted by the red dot evolves to the zero energy state denoted by the black dot, corresponding to evolve to the MKPs as a function of $\mu_{2}$. As an ideal case, the state should remain $|0\tilde{0}\rangle$ during the adiabatic operation. To measure the result, the state formed by the two MKPs should be a bound state. We reverse the above initialization process, localizing the evolved state to the left end of the nanowire to measure the braiding result.  Then, after the total ten steps operations, the initial state $|0\rangle$ denoted by the red dot in Fig.~\ref{fig3}(b) and denoted by  $E_{-1}$ changes to $|1\rangle$ denoted by $E_{1}$ shown by Fig.~\ref{fig3}(e). Due to the time-reversal symmetry, the initial state $|0\tilde{0}\rangle$ changes to $|1\tilde{1}\rangle$ naturally. Since the state $|1\tilde{1}\rangle$ is located at the left end finally, the local measurement can be done to confirm the non-Abelian statistics of MKPs.

\begin{figure}
	\centering
	\includegraphics[width=0.48\textwidth]{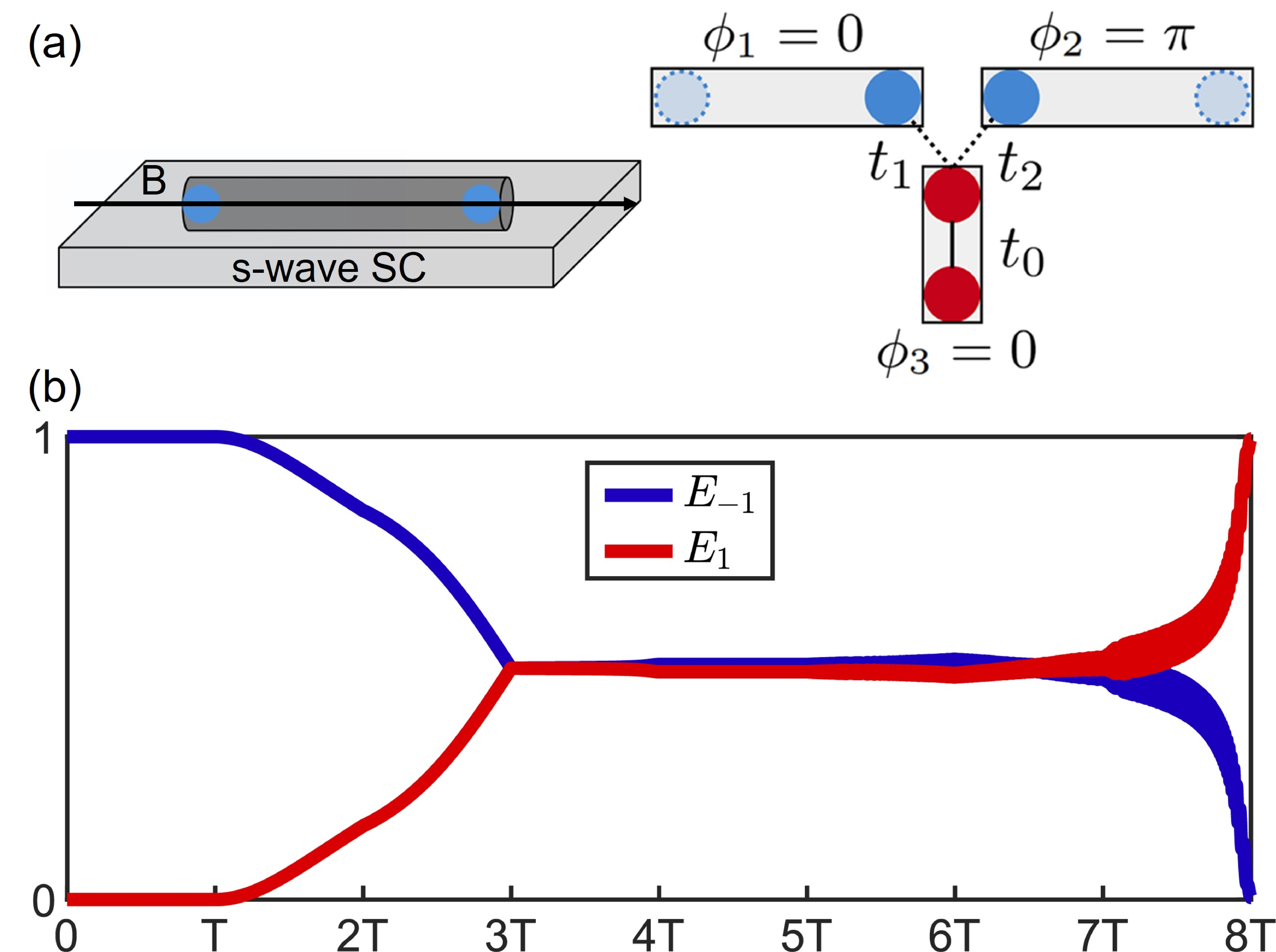}
	\caption{(a) A TSC nanowire (left) and the scheme for braiding (right) of MZM. (b) The evolution of state formed by MZMs in the braiding. $t_{c}$ = 0.6 and T = 2000.}
	\label{fig4}
\end{figure}

\textit{Braiding of MZMs--.} Base on the effect model~\cite{SuppMat}, We can also apply the finite-size effect for the braiding of MZMs~\cite{J. Liu2021}. A nanowire with s-wave paring, Rashba spin-orbit interaction and Zeeman coupling terms can host MZMs at its ends in the topological region~\cite{R. M. Lutchyn2010,Y. Oreg2010}. The Hamiltonian is denoted by
\begin{equation}
	\begin{aligned}
		H_{1 \mathrm{D}} & =-\mu \sum_{j}  c_{j}^{\dagger} c_{j}- t\sum_{j}  c_{j+1}^{\dagger} c_{j}\\
		&+H_{\mathrm{SO}}+H_{\mathrm{SC}}+H_{\mathrm{B}},\\
		H_{\mathrm{SO}} & =-\frac{i}{2} \alpha_R \sum_{j, \alpha, \beta} c_{j+1, \alpha}^{\dagger} \sigma_y^{\alpha, \beta} c_{j, \beta}+ h.c., \\
		H_{\mathrm{SC}} & =\frac{1}{2} \Delta_0 \sum_j c_{j, \uparrow}^{\dagger} c_{j, \downarrow}^{\dagger}+h.c.,\\
		H_{B}&=\sum_{j, \alpha, \beta} c_{j, \alpha}^{\dagger}\left(V_x \vec{\sigma}_x\right)_{\alpha \beta} c_{j, \beta} .
	\end{aligned}
\end{equation}
where $c^{\dagger}_{j}$ ($c_{j}$) creates (annihilates) an electron at site j, $\alpha$ and $\beta$ are the spin indices, $t$ is the nearest
hopping amplitude. The first and second terms are respectively the chemical potential and the kinetic energy; the third term is the spin-orbit coupling with $\alpha_{R}$ being the coupling strength; the fourth term describes the superconducting proximity effect with $\Delta$ being the superconducting s-wave pairing amplitude; the last terms describe the Zeeman couplings with $V_{x}$ being the coupling strength. The braiding scheme is displayed in Fig.~\ref{fig4}(a). To map the effective model, a T-junction is formed by three these nanowires with tunable couplings $t_{1}$ and $t_{2}$. The fixed $t_{0}$ is set by a short nanowire due to the finite-size effect that two MZMs at the ends of the nanowire are coupled. The Hamiltonians of the two long nanowires are denoted by $H_{1}$ and $H_{2}$ with the parameters $N_{1}=100$, $\mu=-2$, $t=1$, $\alpha=0.6$, $\Delta=0.4$, and $V_{x}=1.2$. While the Hamiltonian of the short chain is denoted as $H_{3}$ with $N_{2}=22$ and other parameters being the same as those in $H_{1,2}$. The phases of two long nanowires denoted by $\phi_{1}$ and $\phi_{2}$ are respectively set to be $0$ and $\pi$, while that of the short chain denoted by $\phi_{3}$ is set to $0$, preventing any additional coupling between the MZMs that are not present in the effective braiding model. The coupling Hamiltonians between the two long nanowires and the short nanowire are written as
\begin{eqnarray}
  H_{13}&=&-t_{1}\sum_{\alpha}^{}c^{\dagger}_{1N,\alpha}c_{31,\alpha}+h.c., \nonumber \\
  H_{23}&=&-t_{2}\sum_{\alpha}^{}e^{i\phi_{2}/2}c^{\dagger}_{21,\alpha}c_{31,\alpha}+h.c.,
\end{eqnarray}
where $c^{\dagger}_{1N,\alpha}$ ($c_{1N},\alpha$), $c^{\dagger}_{21,\alpha}$ ($c_{21,\alpha}$) and $c^{\dagger}_{31,\alpha}$ ($c_{31,\alpha}$) are creation (annihilation) operators of electrons at the ends of the nanowires 1, 2, and 3, respectively, which are coupled during the braiding progress. The total Hamiltonian is $H_{0}=H_{1}+H_{2}+H_{3}+H_{13}+H_{23}$. By using the same steps introduced in the effective model, the braiding can be realized. After the braiding, the state $|0\rangle$ denoted by $E_{-1}$ changes to $|1\rangle$ denoted by $E_{1}$ as shown in Fig.~\ref{fig4}(b), corresponding to the transition between the occupied and the non-occupied non-local Fermion states. The method of initialization designed for MKPs can also be used for MZMs.

\textit{Conclusions---.} Based on one-dimensional TSCs, we establish the effective braiding models for MZMs and MKPs in the TSCs without and with time-reversal symmetry, respectively. By utilizing the finite-size effect, we implement the effective model-based braiding in s-wave TSCs, and $d_{x^{2}-y^{2}}$-wave time-reversal symmetric TSCs. Our scheme requires only a few locally tunable coupling parameters and does not require a wide range of moving topological domain walls. Another important problem is the initialization and readout of the state. We design a small gate to control the system to required initial state for the braiding. After braiding, local measurement can be made by reversing the initialization. In particular, the implementation of time-reversal symmetric TSCs does not require strong magnetic fields and avoids the suppression of the resulted superconducting energy gap. Our proposed braiding scheme can be easily utilized to test the non-Abelian statistics of MKPs and MZMs, and therefore implies in the topological quantum computing.

\textit{Acknowledgements---.} We are grateful to Kunhua Zhang for helpful discussions. This work was supported by the National Natural Science Foundation of China (Grants No. 11974327, and 12234017), Anhui Initiative in Quantum Information Technologies (AHY170000), and Innovation Program for Quantum Science and Technology (2021ZD0302800). We also thank the Supercomputing Center of University of Science and Technology of China for providing the high performance computing resources.

\end{document}